# An eco-friendly passivation strategy of resveratrol for highly efficient and antioxidative perovskite solar cells

Xianhu Wu,[a,1] Jieyu Bi,[a,1] Guanglei Cui,[a,*] Nian Liu,[a] Gaojie Xia,[a] Ping Li,[b] Chunyi Zhao,[a] Zewen Zuo,[a] Min Gu[c]

[a] *College of Physics and Electronic Information, Anhui Province Key Laboratory for Control and Applications of Optoelectronic Information Materials, Key Laboratory of Functional Molecular Solids, Anhui Normal University, Wuhu 241002, P. R. China.*

[b] *School of Physics and Electronic Science, Zunyi Normal University, Zunyi 563006, P. R. China*

[c] *National Laboratory of Solid State Microstructures, Nanjing University, Nanjing 210093, P. R. China.*

[1] *These authors contributed equally to this work.*

\* Corresponding Author.

*E-mail address*: glcui@ahnu.edu.cn (G. Cui)

**ABSTRACT**

The stability of perovskite solar cells is closely related to the defects in perovskite crystals, and there are a large number of crystal defects in the perovskite thin films prepared by the solution method, which is not conducive to the commercial production of PSCs. In this study, resveratrol(RES), a green natural antioxidant abundant in knotweed and grape leaves, was introduced into perovskite films to passivate the defect. RES achieves defect passivation by interacting with uncoordinated $Pb^{2+}$ in perovskite films. The results show that the quality of the perovskite film is significantly improved, and the energy level structure of the device is optimized, and the power conversion efficiency of the device is increased from 21.62% to 23.44%. In addition, RES can hinder the degradation of perovskite structures by $O_2^-$ and $CO_2^-$ free radicals, and the device retained 88% of its initial PCE after over 1000 hours in pure oxygen environment. The device retains 91% of the initial PCE after more than 1000 hours at 25°C and 50±5% relative humidity. This work provides a strategy for the use of natural and environmentally friendly additives to improve the efficiency and stability of devices, and provides an idea for the development of efficient, stable and environmentally friendly PSCs.

# 1. Introduction

In the past decade, significant advancements have been made in the research of organic-inorganic hybrid perovskite solar cells (PSCs), with their certified power conversion efficiency (PCE) exceeding 26%, surpassing those of solar cells fabricated from polycrystalline and monocrystalline silicon [1-5]. Due to their advantages such as facile solution processing and low-temperature fabrication, [6-8] they also hold great potential for widespread applications in areas including LEDs, photodetectors, and sensors [9-12]. However, the poor stability of perovskite crystals remains one of the major challenges facing PSCs, necessitating resolution before practical deployment.

According to previous studies, the stability of PSCs in pure oxygen environments is poor, mainly because $O_2$ can weakly interact with perovskite, allowing excess electrons generated by light excitation in perovskite to transfer to the surface of perovskite, where they form superoxides and peroxides with $O_2$, leading to the disintegration of the Pb-I octahedral structure of perovskite. [13,14] Additionally, under light exposure, $O_2$ and $CO_2$ can acquire electrons carried by $I^-$ in perovskite to form $O_2^-$ and $CO_2^-$ radicals, which break down $FA^+$ and $MA^+$ groups in perovskite to obtain protons, forming $H_2O$ and $HCO_2$. [15] This irreversible process disrupts the octahedral structure of perovskite. Furthermore, during the growth of perovskite crystals, a large number of defects are formed at grain boundarie and surface, which act as non-radiative recombination centers, inhibiting charge carrier transport and promoting perovskite decomposition. [16] To reduce defects and improve the stability of perovskite, many molecular materials are used for doping perovskite. Among them, NaDTE with C=O groups is introduced to passivate uncoordinated lead ions in perovskite, significantly reducing the defect density of perovskite. [17] The phosphonic acid group in dimethylamine is used to coordinate with lead iodide in perovskite, leading to p-type doping of the perovskite film. [18] The C-N and C=N groups in 1H-benzimidazole (BIE-H) and 1-(6-bromohexyl)-1H-benzimidazole (BIE-Br) are used to passivate uncoordinated $Pb^{2+}$ defects in perovskite through Lewis acid-base coordination, suppressing non-radiative recombination of charge carriers. [19] Organic molecule phenylphosphonic acid (PPA) is used for doping perovskite, and the P=O group in PPA has a strong coordinating interaction with uncoordinated $Pb^{2+}$, which helps to reduce the defect density. [20] The hydroxamic acid group in benzohydroxamic acid (BHA) not only alleviates the oxidation of $Sn^{2+}$ but also interacts with metal ions at the B site to passivate relevant defects. [21] However, most of the additives currently used have complex synthetic processes and high prices. Therefore, it is necessary to find inexpensive, simple, and environmentally friendly molecular materials in nature. The strong interaction between caffeine and uncoordinated $Pb^{2+}$ ions in perovskite can act as a "molecular lock," increasing the activation energy during film crystallization and reducing ion migration. [22] Vanillin introduced into perovskite precursors can inhibit the oxidation of $Sn^{2+}$ in perovskite. [23] Lycopene has conjugated double bonds. It introduced into perovskite precursors can act as a Lewis base additive to passivate uncoordinated $Pb^{2+}$. [24] Cocoa alkaloids is introduced into perovskite precursors, the hydrogen bonds between N-H and I help to combine C=O with opposite Pb defects, thus maximizing defect passivation. [25] Natural vitamin B can simultaneously passivate positively and negatively charged ion defects in perovskite, which is beneficial for charge transport in devices. [26] Theine can increase the formation energy of most defect on the surface of perovskite, thereby inhibiting defect generation. It can also effectively eliminate or weaken defects in the bandgap. [27] Therefore, finding suitable natural and environmentally friendly molecular materials in nature is significant for improving the stability and efficiency of PSCs. Resveratrol, a non-flavonoid polyphenolic organic compound found in nature, has strong antioxidant properties and can be synthesized through a simple process in polygonum cuspidatum and grape leaves.

The antioxidative compound resveratrol is employed for doping perovskite to passivate grain boundary defects and regulate the growth of perovskite crystals. The C=C bond and the oxygen atom in the -OH group of resveratrol, which carries a negative charge, harbor lone pair electrons, allowing them to function as Lewis bases for coordinating with uncoordinated $Pb^{2+}$ ions within the perovskite lattice and passivating $I^-$ defects in the perovskite structure. Additionally, the -OH moiety (phenolic hydroxyl group) is capable of scavenging $O_2^-$ and $CO_2^-$ radicals, hindering the degradation of $FA^+$ and $MA^+$ groups, thus impeding and retarding the crystalline degradation of perovskite, consequently enhancing the stability of PSCs. Ultimately, devices based on resveratrol-doped perovskite achieved a PCE of 23.44%, with unencapsulated devices retaining 90% of their initial efficiency after exposure to an environment with 55±5% relative humidity (RH) and 25°C for over 1000 hours. This green additive is conduct to advancing the commercial production of PSCs.

## 2. Experimental

*2.1. Materials*

FABr (99.5%), 2,2′,7,7′-tetrakis(N,Ndi- p-methoxyphenylamine)-9,9′-spirobifluorene (Spiro-OMeTAD) (99.5%), Formamidine Hydroiodide (FAI, 99.99%), lead(II) iodide (PbI$_2$, 99.99%), LiTFSI (99%), CsI (99.99%) and bathocuproine (BCP) were purchased from *Xi'an Yuri Solar Co., Ltd*. The SnO$_2$ colloidal dispersion (SnO$_2$, 15%) was purchased from *Alfa Aesar*. PC61BM, chlorobenzene (CB) and etched indium tin oxide (ITO) substrates are all purchased from *Advanced Election Technology Co., Ltd*. Dimethyl sulfoxide (DMSO, 99.9%), 4-Tert-Butylpyridine (TBP, 96%) and N,N-Dimethylformamide (DMF, 99.9%) were purchased from *Sigma-Aldrich*. Resveratrol (3-4'-5-trihydroxystilbene) were purchased from *SHANDONG SHENGJIADE BIOTECHNOLOGY Co., Ltd.*

*2.2. Sample preparation*

The ITO substrates were ultrasonically cleaned with deionized water, isopropanol, and absolute ethanol for 15 minutes, respectively, and then the cleaned ITO substrates were blown dry with nitrogen. Then take 40 μL of SnO$_2$ solution drops on ITO and spin coat 4000 rpm/min for 30 s and anneal on a heating plate at 150 °C for 30 min to form a dense layer of SnO$_2$. Then the substrate is plasma-treated for 20 minutes. The control and Resveratrol doped perovskite films were spin-coated at 5000 rpm/min for 30 s, and 200 μL of CB was dropped vertically on the perovskite at 24 s as an anti-solvent, followed by annealing at 150 °C for 10 min. Detailed preparation processes can be found in supplementary material.

**3. Result and discussion**

*3.1 Effect of RES doped perovskite on the surface properties and optical properties of*

*CSFAMA thin films*

The simple experimental procedure for doping perovskite (CSFAMA) with resveratrol (RES) is illustrated in **Fig. 1** (detailed experimental procedures are provided in the supplementary material). By introducing varying concentrations of RES, modifications were made to the crystal facets and grain boundaries of the perovskite, followed by a systematic evaluation of the CSFAMA thin films and their device performance. The optimal concentration of RES was determined to be 0.2 wt%, and this concentration was employed for the analyses conducted in this study. (Unless otherwise specified, the RES discussed hereafter refers to 0.2 wt% RES).

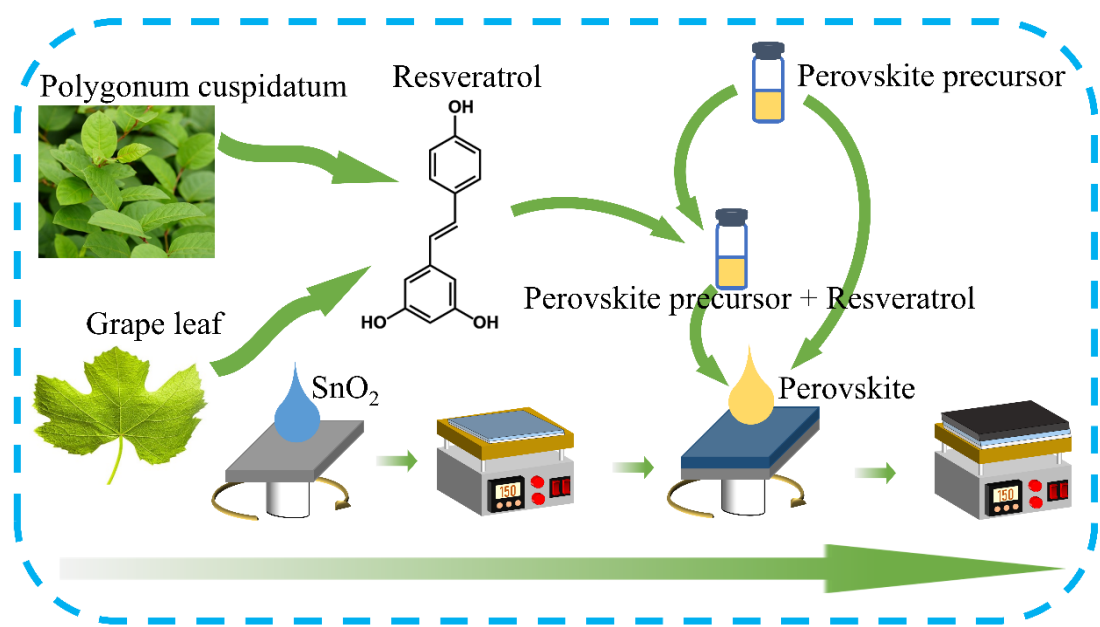

**Fig. 1**. Schematic diagrams of the preparation of CSFAMA and RES-CSFAMA.

To investigate the influence of RES doping on the morphology of CsFAMA, scanning electron microscope (SEM) images of undoped and RES-doped CsFAMA thin films were measured and presented in **Fig. 2a and 2b**. It can be observed that the grains of both undoped and RES-doped CsFAMA thin films are dense, but the surface of undoped CsFAMA thin film exhibits more PbI$_2$ compared to RES-doped CsFAMA thin film, indicating the passivation of PbI$_2$ by RES. The average grain sizes of perovskite grains within the ranges depicted in **Fig. 2a and 2b** were respectively calculated and shown in **Fig. S1a and S1b**, with average grain sizes of 0.46 and 0.81 μm for undoped and RES-doped CsFAMA thin films, respectively. The larger grain size of perovskite thin films doped with RES suggests that RES influences the crystallization and nucleation process of perovskite, facilitating the enlargement of perovskite grains. The increase in grain size is advantageous for reducing the interfacial area, thereby reducing non-radiative recombination centers within the perovskite bulk and on the film surface. [28] Cross-sectional SEM images of undoped and RES-doped perovskite thin films were measured and presented in **Fig. S1c and S1d**, with thicknesses around 650 nm for both undoped and RES-doped CsFAMA thin films. Although the thickness of CsFAMA remains unchanged, in undoped CsFAMA, grains are disordered and stacked, leading to dangling bonds or vacancies at grain boundaries, which serve as non-radiative recombination centers. In contrast, grains in RES-doped CsFAMA are more regular, with fewer non-radiative recombination centers, facilitating charge transport. [29] Atomic force microscopy (AFM) images of undoped and RES-doped CsFAMA thin films were measured and presented in **Fig. 2c and 2d**. Compared to CsFAMA thin film, the surface roughness of RES-doped CsFAMA thin film is significantly reduced from 31.8 nm to 23.1 nm, indicating that RES additive contributes to the formation of smooth and dense CsFAMA thin films. X-ray diffraction (XRD) spectra of undoped and RES-doped CsFAMA thin films are shown in **Fig. 2e**. Compared to CsFAMA, the diffraction peaks of RES-doped CsFAMA thin films show no significant changes, indicating that RES molecules are located at the grain boundaries of CsFAMA without causing changes to the CsFAMA lattice, enhancing the crystalline quality of RES-doped CsFAMA thin films. [17,30] Optical absorption spectra of undoped and RES-doped CsFAMA thin films are presented in **Fig. 2f**. Compared to CsFAMA thin film, RES-doped CsFAMA thin film exhibits significantly enhanced light absorption below 580 nm, which is advantageous for increasing the number of photogenerated carriers in perovskite, thereby enhancing the device's Jsc. Tauc plots of undoped and RES-doped CsFAMA thin films demonstrate that the bandgap of perovskite is 1.525 eV, further indicating that RES molecules do not dope into the lattice of perovskite but rather reside at the grain boundaries, promoting the growth of perovskite grains and passivation of defects at grain boundaries. [17,29-31]

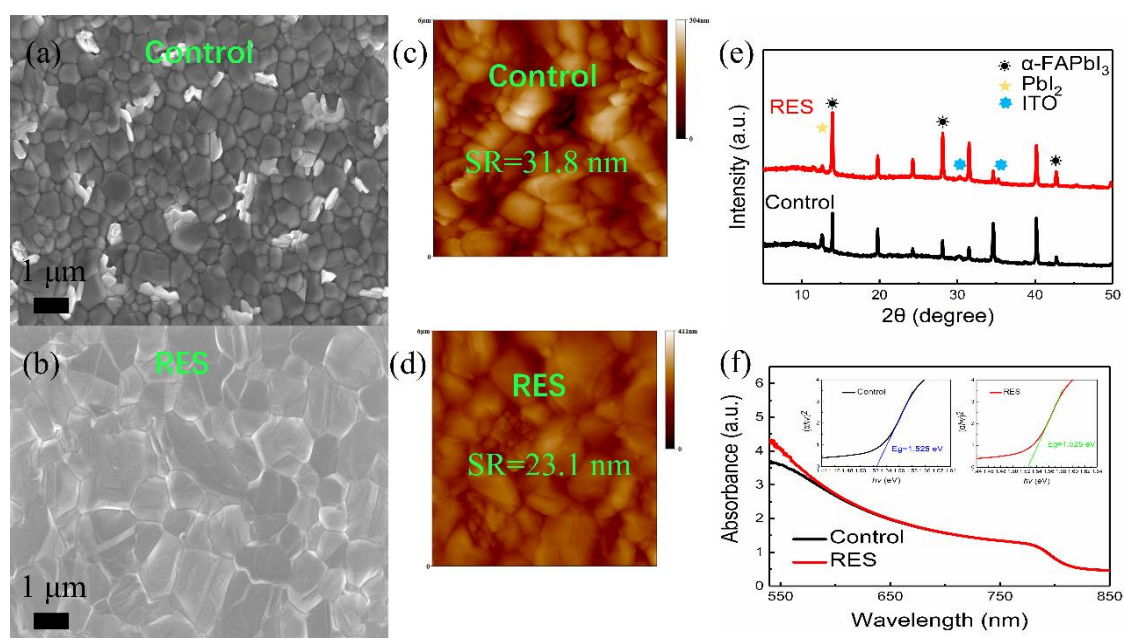

**Fig. 2**. SEM images of (a) CSFAMA and (b) RES-CSFAMA thin films. AFM images of (c) CSFAMA and (d) RES-CSFAMA thin films. The XRD of CSFAMA and RES-CSFAMA thin films. (f) The optical absorption spectra of

CSFAMA and RES-CSFAMA thin films.

To study the effect of RES-doped perovskites on the energy level structure of perovskites, the ultraviolet photoelectron spectra of undoped and RES-doped perovskite films were measured and shown in **Fig. S2**, and the cut-off binding energies of the undoped and RES-doped perovskite films were 16.76 and 16.87 eV, respectively. The work functions of the undoped and RES-doped CSFAMA films were -4.46 and -4.35 eV, respectively, by the formula $E_F = E_{cut-off} - 21.22\ eV$. The Fermi edges of undoped and RES-doped CSFAMA films were 1.42 and 1.47 eV, respectively. According to the formula $E_{VB} = E_F - E_{F,edge}$, the valence band tops (VBM) of the undoped and RES-doped CSFAMA films were -5.88 and -5.82 eV, respectively. Based on the optical band gap of the undoped and RES-doped perovskite films in **Fig. 2f** (1.525 eV rounded to 1.53 eV), the conduction band bottoms (CBM) of the undoped and RES-doped perovskite films are calculated to be -4.35 and -4.29 eV, respectively. Based on the above calculation results, the energy level structures of undoped and RES-doped CSFAMA films are obtained and shown in **Fig. 3a**. The RES doping perovskite shifts the Fermi level and valence band top of CSFAMA upward, which is conducive to promoting the transport of holes at the interface of perovskite layer/hole transport layer. To investigate the passivation effect of RES on perovskites, UV photoelectron spectroscopy (XPS) of undoped and RES-doped CSFAMA films were measured, as shown in **Fig. 3b and 3c**. After RES doping, the binding energies of Pb 4f and I 3d changed, indicating that there was an interaction between RES and perovskites. In the control sample, the two main peaks of $Pb^{2+}$ 4f were located at 138.1 and 142.9 eV. In

RES-doped perovskites, the two main peaks of $Pb^{2+}$ are shifted by 0.21 eV towards the low binding energy. The behavior of this binding energy shifting in a lower direction indicates that there is a strong interaction between RES and $Pb^{2+}$. In the control sample, there were two small peaks attached to the $Pb^{2+}$ peak at 136.6 and 141.1 eV, which were uncoordinated defects of the metal $Pb^0$. This defect causes the formation of a complex center of deep energy level defects in the perovskite, which directly traps the carriers in the perovskite layer. [32] In the RES-doped CSFAMA film, the $Pb^0$ defect disappeared completely, indicating that RES could interact with the uncoordinated $Pb^{2+}$ in CSFAMA. The passivation effect of RES is further illustrated by the electrostatic potential (ESP) of the RES molecule in **Fig. 3d**. The -OH functional group in RES shows high electron density and negative potential in the blue area of the ESP plot. The O in -OH contains lone pairs, and the lone pairs of C=C double bonds can be used as Lewis bases. [24,29,33] Whereas, uncoordinated $Pb^{2+}$ is a Lewis acid. According to Lewis theory, uncoordinated $Pb^{2+}$ can accept electrons in the RES molecule to form Lewis acid-base coordination with C=C and -OH through covalent coordination bonds. [17,24] This is beneficial to promote the degradation of $Pb^0$-induced deep level defects, thereby reducing the non-radiative recombination center in CSFAMA and improving the performance of perovskite films and devices. **Fig. 3e** illustrates a clear schematic of the interaction between RES and perovskite grain boundaries.

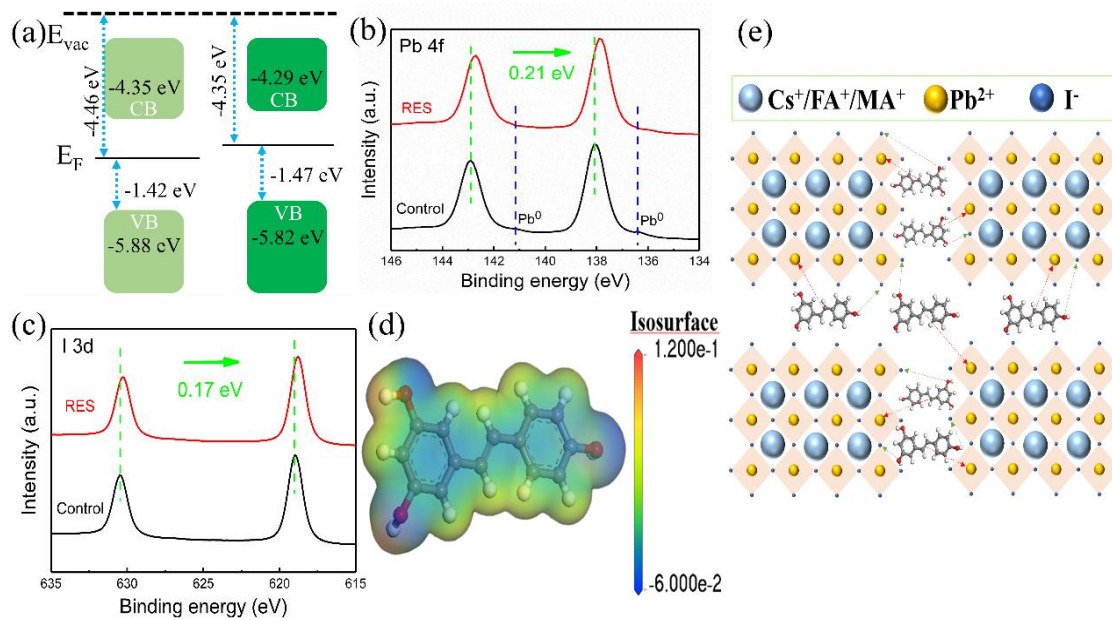

**Fig. 3**. (a) The energy level structure of CSFAMA and RES-CSFAMA thin films. (b) The Pb 4f and (c) I 3d peaks of CSFAMA and RES-CSFAMA thin films. (d) ESP distribution of RES molecule. (e) Schematic diagram of RES interacts with perovskite.

*3.2 Carrier dynamics of the interface between RES-CSFAMA/PEAI/Spiro-OMETAD*

To investigate the effect of RES doping on the optical properties of perovskites, the steady-state PL spectra of CSFAMA films prepared on pure glass substrates are shown in **Fig. 4a**. Compared with the undoped CSFAMA, the steady-state PL peak intensity of the CSFAMA sample doped with RES was significantly enhanced, indicating that the CSFAMA doped with RES could produce more photogenerated carriers under the same illumination. To further investigate the carrier dynamics of the interface (the interface between CSFAMA and PEAI/Spiro-OMeTAD), we measured the steady-state PL of CSFAMA/PEAI/Spiro-OMeTAD and RES-CSFAMA/PEAI/Spiro-OMeTAD and shown in **Fig. S3**. The RES-based samples exhibit significant PL quenching, which directly proves that holes are easier to extract at the interface. To investigate the effect of RES doping CSFAMA on the transient carrier dynamics of the CSFAMA/PEAI/Spiro-OMeTAD interface. We used a 450 nm (5 μJ cm$^{-2}$) femtosecond laser to excite the samples and measure the ultrafast femtosecond transient absorption spectra of CSFAMA/PEAI/Spiro-OMeTAD and RES-CSFAMA/PEAI/Spiro-OMeTAD (**Fig. 4b and 4c**). It can be seen that the ground state bleaching peak of both groups of samples is 785 nm, and the bleaching attenuation of RES-CSFAMA/PEAI/Spiro-OMeTAD is more significant than that of CSFAMA/PEAI/Spiro-OMeTAD. We then extracted the ground state bleaching kinetics of CSFAMA/PEAI/Spiro-OMeTAD and RES-CSFAMA/PEAI/Spiro-OMeTAD, respectively (Fig. 4d), and the ground state bleaching kinetics of RES-CSFAMA/PEAI/Spiro-OMeTAD decayed significantly faster than that of CSFAMA/Spiro-OMeTAD, which was attributed to the modified CSFAMA/PEAI/ Superior hole extraction rate at the Spiro-OMeTAD interface. In order to quantitatively illustrate the difference in charge transport, the ground state bleaching kinetics were fitted, and the results are shown in **Table S1**, where $\tau_1$, $\tau_2$ and $\tau_3$ are the hole transfer time, the electron-hole pair recombination time, and the defect capture time, respectively [34]. By calculation, we can obtain that the hole transfer rates are $7.2 \times 10^9$ s$^{-1}$ (CSFAMA/PEAI/Spiro-OMeTAD) and $4.1 \times 10^{10}$ s$^{-1}$ (RES-CSFAMA/PEAI/Spiro-OMeTAD), respectively, which is consistent with the difference in PL intensities observed previously (**Fig. S3**), which again illustrates the effect of RES doping CSFAMA on CSFAMA/PEAI/ The extraction efficiency of Spiro-OMeTAD interfacial holes was significantly higher than that of undoped CSFAMA. This is beneficial to improve the photovoltaic performance of PSCs.

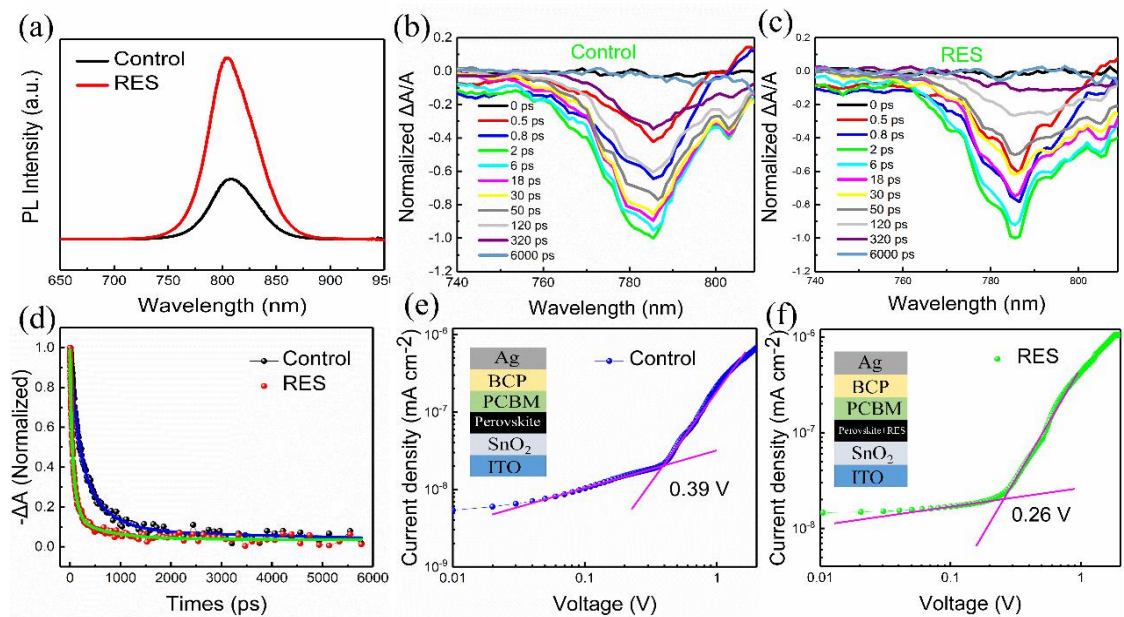

**Fig. 4.** (a) PL of CSFAMA and RES-CSFAMA films. The ultrafast transient absorption spectra of (b) CSFAMA/PEAI/Spiro-OMeTAD and (c) RES-CSFAMA/PEAI/Spiro-OMeTAD. (d) The GSB dynamics of CSFAMA/PEAI/Spiro-OMeTAD and RES-CSFAMA/PEAI/Spiro-OMeTAD. SCLC measurements of electron-only (e) CSFAMA and (f) RES-CSFAMA.

The space charge limited current (SCLC) was tested and depicted in **Fig. 4e and 4f**. The J-V curve of pure electron devices can be divided into three regions: ohmic region, defect-filled region with sharply increasing current, and defect-free region. [35] After RES modification, $V_{TFL}$ decreased from 0.39 to 0.26 V, and then the trap density of the CSFAMA thin film was calculated using equation [51]: $N_{traps} = 2\varepsilon_0\varepsilon V_{TFL}/eL^2$, where $\varepsilon$ is the relative dielectric constant of perovskite, $\varepsilon_0$ is the vacuum permittivity, $e$ is the charge, and $L$ is the thickness of the CSFAMA film (approximately 650 nm as shown in **Fig. S1**). After RES doping, the calculated defect density significantly decreased from $3.14\times10^{15}$ cm$^{-3}$ to $2.09\times10^{15}$ cm$^{-3}$ (**Table S2**). The above results indicate that the C=C and -OH groups in the RES molecule significantly reduce the defects in the perovskite film by coordinating with $Pb^{2+}$ in CSFAMA. Subsequently, electrochemical impedance spectroscopy (EIS) was also employed to further understand the interface charge transfer and recombination. The Nyquist plots of the devices with undoped and RES-doped CSFAMA obtained through EIS measurements are shown in **Fig. S4**. The fitted EIS spectra results, including series resistance ($R_s$), recombination resistance ($R_{rec}$), and transport resistance ($R_{tr}$), are listed in Table S3. After RES doping, $R_{tr}$ decreased from 354.2 Ω to 326.7 Ω, indicating an improvement in charge transfer and extraction in the devices. $R_{rec}$ increased from 1404 to 1912 Ω, indicating effective suppression of defects in the perovskite by RES doping. The reduction in defects is mainly attributed to the coordination of C=C and -OH groups in RES with uncoordinated $Pb^{2+}$ in the perovskite.

*3.3 Effect of RES-doped CSFAMA on photovoltaic performance of perovskite solar cells*

Under standard simulated solar irradiation conditions (AM 1.5 G, 100 mW $cm^{-2}$), the J-V characteristics of PSCs based on undoped and RES-doped CSFAMA thin films are depicted in **Fig. 5a**. The optimal $V_{oc}$, $J_{sc}$, FF, and PCE of the devices increase from 1.121 V, 25.24 mA $cm^{-2}$, 76.44%, and 21.62% for the control device to 1.154 V, 25.72 mA $cm^{-2}$, 78.95%, and 23.44% for the RES-doped device, respectively. The performance of the devices is significantly enhanced. As illustrated in **Fig. 5b and Table S4**, the RES-based devices exhibit excellent reproducibility. The increase in $V_{oc}$ is attributed to better energy level alignment at the perovskite/PEAI/Spiro-OMeTAD interface, while the increase in Jsc mainly originates from superior perovskite crystallization, leading to enhanced light absorption and improved charge carrier transport. The EQE spectra of the control and RES-based devices are shown in **Fig. 5c**. The integrated current density increases from 24.51 to 25.18 mA cm$^{-2}$, consistent with the trend observed in the J-V curves. Furthermore, the devices after RES doping exhibit weaker hysteresis (hysteresis index, HI=0.04) compared to the control devices (HI=0.09) (**Fig. 5d, 5e,** and **Table S5**), indicating better suppression of hysteresis effects in the RES-doped devices. Additionally, we conducted stability tests at the maximum power point for 240 seconds for both control and RES-doped devices (**Fig. 5f**). After 240 s, the current density and PCE of the control device are 24.354±1.119 mA cm$^{-2}$ and 20.424±1.151%, respectively, while those of the RES-doped device are 25.155±0.276 mA cm$^{-2}$ and 22.958±0.325%, respectively. In contrast, the performance of the RES-doped devices remains highly stable over the same duration.

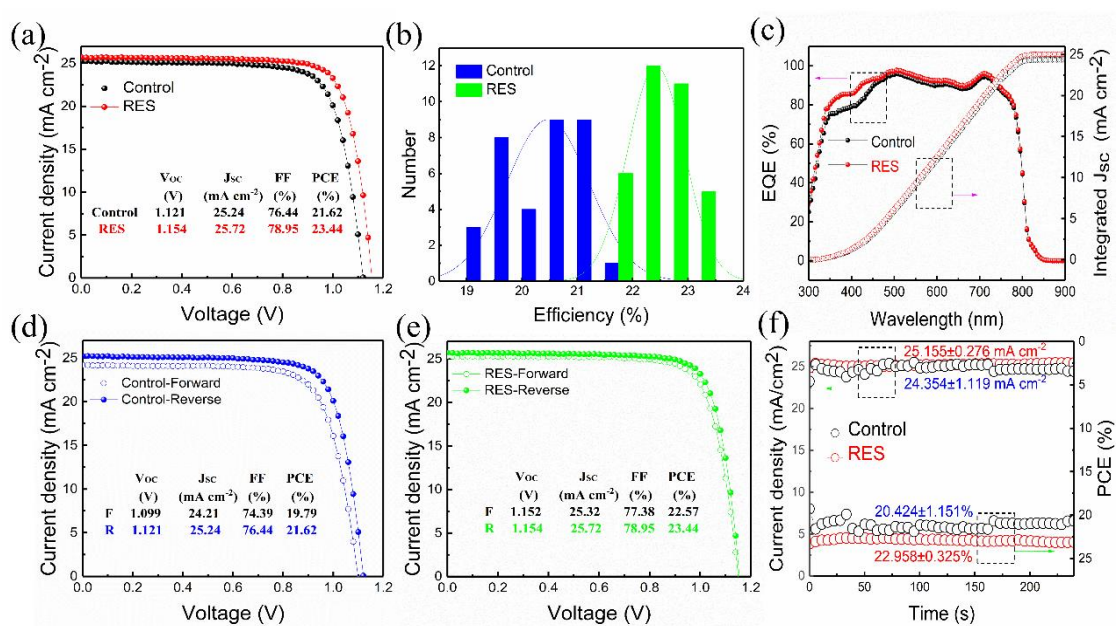

**Fig. 5**. (a) J-V curves. (b) Efficiency distribution histogram. (c) EQE spectra. The J-V curves of devices based on (d) CSFAMA and (e) RES-CSFAMA films in different scanning directions. (f) Stabilized output current densities and efficiencies at maximum power points of the optimized devices based on CSFAMA and RES-CSFAMA films.

Stability is crucial for devices alongside their photovoltaic performance. Therefore, long-term stability tests were conducted on unpackaged devices under dark conditions (humidity approximately 50±5%, 25°C). As depicted in **Fig. 6**, after approximately 1000 hours of storage, devices based on CAFAMA retained only 66% of their initial PCE, while those based on RES retained 91% of their initial PCE. The improved environmental stability of the devices can be attributed to the increase in water contact angle from 58% to 75% after RES doping of the CSFAMA film (**Fig. S5**). Stability measurements in pure $O_2$ environment at room temperature are shown in **Fig. S6**. Due to the excellent antioxidative performance of RES, the devices exhibited an 88% retention of initial PCE after 1000 hours, compared to 61% retention for the control devices. Under illumination, the electrons carried by I- in the perovskite transfer to $O_2$ and $CO_2$ to form $O_2^-$ and $CO_2^-$ radicals, which then attack $FA^+$ and $MA^+$ groups to acquire protons, forming $H_2O$ and $HCO_2$. [15] This irreversible destruction of the perovskite crystal structure occurs. The enhanced stability of the devices in a pure $O_2$ environment can be attributed to the presence of -OH groups (phenolic hydroxyls) in RES, which can bind with $O_2^-$ and $CO_2^-$ radicals in the aforementioned process, thereby hindering and delaying the degradation of $FA^+$ and $MA^+$ groups by $O_2^-$ and $CO_2^-$ radicals, enhancing the antioxidative properties of the CSFAMA film, and consequently improving the stability of the perovskite crystal structure and the device.

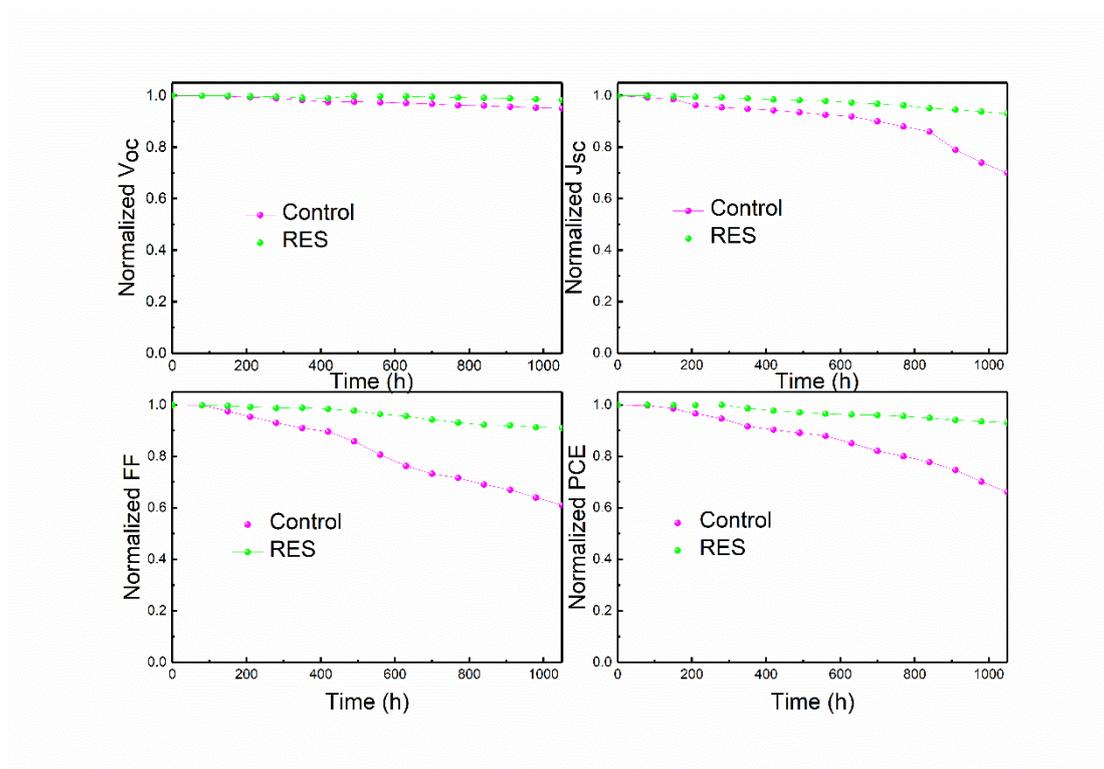

**Fig. 6.** Stability of Voc, Jsc, PCE and FF of the unencapsulated PSCs based on the CSFAMA and CSFAMA films in 50 ± 5% RH ambient air.

3. Conclusions

In summary, the naturally green antioxidant, resveratrol (RES), extracted from tiger cane and grape leaves, has conjugated double bonds and phenolic hydroxyl groups, facilitates the enhancement of the stability of perovskite solar cells (PSCs). Through Lewis acid-base coordination between C=C and $Pb^{2+}$, RES effectively inhibits grain boundary defects in perovskite films and improves film crystallinity. Ultimately, the PCE of PSCs is increased from 21.62% to 23.44%. Due to the strong antioxidant properties of phenolic hydroxyl groups, RES effectively eliminates $O_2^-$ and $CO_2^-$ radicals, efficiently hinder and mitigate the damage of radicals to the perovskite crystal structure. The introduction of RES additives imparts perovskite with pronounced hydrophobicity and antioxidative properties. It enables the device to retain 88% of its initial PCE after 1000 hours in pure $O_2$ environment, and preserving 91% of the initial PCE after over 1000 hours at 25°C and 50±5% relative humidity. Our study underscores the importance of selecting suitable molecular materials with green natural antioxidant properties from nature to manufacture efficient and stable PSCs, providing researchers with a perspective for exploring natural materials for the fabrication of efficient and stable PSCs.

**Conflicts of interest**

There are no conflicts to declare.

**Acknowledgements**

This work was supported by the National Natural Science Foundation of China through Grants 21373011, 12264060 and 51871003, Anhui Provincial Natural Science Foundation (2108085MA24), and the University Synergy Innovation Program of Anhui Province (GXXT-2021-049). This work was also supported by the Scientific Research Foundation of Guizhou Province Education Ministry (Grant No. QJHKYZ[2020]037).